\documentclass{article}
\usepackage{amsmath}
\usepackage{cases}
\usepackage{url}
\usepackage[inline]{enumitem}
\usepackage{threeparttable}

\begin{document}
\title{Cold-start recommendations in Collective Matrix Factorization}
\author{David Cortes}
\maketitle

\begin{abstract}
This work explores the ability of collective matrix factorization models in recommender systems to make predictions about users and items for which there is side information available but no feedback or interactions data, and proposes a new formulation with a faster cold-start prediction formula that can be used in real-time systems. While these cold-start recommendations are not as good as warm-start ones, they were found to be of better quality than non-personalized recommendations, and predictions about new users were found to be more reliable than those about new items. The formulation proposed here resulted in improved cold-start recommendations in many scenarios, at the expense of worse warm-start ones.
\end{abstract}

\section{Introduction}

This work aims to explore the quality of cold-start recommendations derived from collective matrix factorization models \cite{cmf} in collaborative filtering with explicit-feedback data in the form of ratings. Recommender systems based on collaborative filtering are typically constructed solely based on data about user-item interactions \cite{koren}, such as movies rated by different users, which result in domain-independent and easily-implementable models, but have the disadvantage of only being able to make recommendations about users and items for which there is interactions data available (known as \emph{warm-start} recommendations in the literature).

In many settings however, there is oftentimes additional side information available about users and/or items, which is not used in the most common models such as low-rank matrix factorization \cite{koren} or kNN-based formulas \cite{konstan}, but which can be used both to improve recommendation models that take interactions data, and to make recommendations in the absence of interactions data (so-called \emph{cold-start} recommendations).

This work focuses on the second case: studying recommendations from matrix factorization models that are based on attributes data without interactions data.

\section{Collective Matrix Factorization}

Collective matrix factorization is an extension of the low-rank factorization model that tries to incorporate attributes about the users and/or items by also factorizing the matrices associated with their side information, sharing the latent factors between them.

More formally, recommendation models based on low-rank matrix factorization try to factorize a partially-observed matrix $\mathbf{X}_{ui}$ of user-item interactions (e.g. movie ratings), where $u$ is the number of users and $i$ is the number of items, into the product of two lower-dimensional matrices $\mathbf{A}_{uk}$ and $\mathbf{B}_{ik}$, where $k \ll u, i$, which can be thought of as latent factors determined for each user and item, by minimizing some loss function, such as squared loss, defined only on the entries of $\mathbf{X}_{ui}$ that are known (hereafter denoted by the indicator function $I_x$), e.g.:

$$ \min_{\mathbf{A},\mathbf{B}} \|I_x(\mathbf{X} - \mathbf{A} \mathbf{B}^T)\| $$

Having obtained these matrices, it’s then possible to predict the values of $\mathbf{X}$ for entries that are not known by the dot product $<\mathbf{a}_u, \mathbf{b}_i>$ for user $u$ and item $i$. Recommendations are then made by sorting these predictions in descending order.

In most implementations, this model is improved by centering the data (subtracting the global mean $\mu$ from each entry), adding user and item biases $\mathbf{m}_u$ (row vector) and $\mathbf{n}_i$ (column vector), which might be treated as model parameters or obtained through a simple heuristic before attempting to obtain optimal values for $\mathbf{A}$ and $\mathbf{B}$, and by adding regularization on all the model parameters, resulting in the following problem:

$$\min_{\mathbf{A},\mathbf{B}} \| I_x( \mathbf{X} - \mu - \mathbf{m} - \mathbf{n} - \mathbf{A} \mathbf{B}^T) \|^2  + \lambda( \| \mathbf{A} \|^2 + \| \mathbf{B} \|^2) $$

This is a non-convex optimization problem for which local minima can be found either by gradient-based methods, or more typically, by the ALS (alternating least-squares) algorithm \cite{koren}\cite{als}, which takes advantage of the fact that, if holding one of the low-rank matrices constant, the optimal values for the other can be obtained through a closed-form solution that implies solving linear systems – the algorithm then alternates between solving one or the other holding the other constant until convergence.

The main idea behind collective matrix factorization is to jointly factorize the interactions matrix $\mathbf{X}_{ui}$ along with the user attributes matrix $\mathbf{U}_{up}$ and the item attributes matrix $\mathbf{I}_{iq}$, introducing new matrices $\mathbf{C}_{pk}$ and $\mathbf{D}_{qk}$ for the user and item attributes (assuming there is data about both user and item attributes), but sharing the $\mathbf{A}_{uk}$ and $\mathbf{B}_{ik}$ matrices between factorizations:

$$ \min_{\mathbf{A},\mathbf{B},\mathbf{C},\mathbf{D}} \|I_x (\mathbf{X} - \mathbf{A} \mathbf{B}^T) \|^2 + \| \mathbf{U} - \mathbf{A} \mathbf{C}^T \|^2 + \| \mathbf{I} - \mathbf{B} \mathbf{D}^T \|^2 $$

Up to this point, the problem is equivalent to factorizing an extended block matrix $ \mathbf{X}_{ext} = 
  \left(\begin{array}{@{}cc@{}}
    \mathbf{X} & \mathbf{U} \\
    \mathbf{I}^T & .
  \end{array}\right)
$ and can be solved using the same methods as before.

The new matrices $\mathbf{C}_{pk}$ and $\mathbf{D}_{qk}$ are not used in the prediction formula, but their presence in the minimization objective allows obtaining better estimates for $\mathbf{A}$ and $\mathbf{B}$ - informally, they now need to explain both the interactions and the side information, making them less prone to overfitting the observed interactions and forcing these latent factors to relate to the non-latent attributes, thereby generalizing better to new data.

There are many logical improvements upon this model: the matrices might not share all the latent factors, but have independent parts, e.g.
$$\mathbf{A} = \left(\begin{array}{@{}ccc@{}} \mathbf{A}_{\text{attr}} & \mathbf{A}_{\text{shared}} & \mathbf{A}_{\text{main}} \end{array}  \right) , \: \: \:
\mathbf{B} = \left(\begin{array}{@{}ccc@{}} \mathbf{B}_{\text{attr}} & \mathbf{B}_{\text{shared}} & \mathbf{B}_{\text{main}} \end{array}  \right)$$
each factorization might have a different weight, each matrix its own regularization hyperparameter, among others. Particularly, this work also applied a sigmoid transformation to all binary variables in the side information matrices, took the user and item biases as model parameters, for which regularization was also applied, and divided the sum of residuals from each matrix by the number of entries in order for their contribution not to be driven by the relative size of each, resulting in an optimization problem as follows:
$$
\min_{\mathbf{A},\mathbf{B},\mathbf{C},\mathbf{D},\mathbf{m},\mathbf{n}} \: \: \:
\mathrm{w}_x \mathrm{L}_x(\mathbf{X}, \mathbf{A}_x, \mathbf{B}_x) +
\mathrm{w}_u \mathrm{L}_u(\mathbf{U}, \mathbf{A}_u, \mathbf{C}) +
\mathrm{w}_i \mathrm{L}_i(\mathbf{I}, \mathbf{B}_i, \mathbf{D}) +
\mathrm{R}
$$
where:\\

$\mathrm{L}_x(\mathbf{X}, \mathbf{A}_x, \mathbf{B}_x) =
\frac{\| I_x( \mathbf{X} - \mu - \mathbf{m} - \mathbf{n} - \mathbf{A}_x \mathbf{B}_x^T) \|^2}{|\mathbf{X}|} $

$\mathrm{L}_u(\mathbf{U}, \mathbf{A}_u, \mathbf{C}) =
\frac{\| \mathbf{U} - \mathrm{S}(\mathbf{A}_u \mathbf{C}^T) \|^2}{|\mathbf{U}|} \: \: , \: \:
\mathrm{L}_i(\mathbf{I}, \mathbf{B}_i, \mathbf{D}) = 
\frac{\| \mathbf{I} - \mathrm{S}(\mathbf{B}_i \mathbf{D}^T) \|^2}{|\mathbf{I}|} $

$\mathrm{R}(\mathbf{A},\mathbf{B},\mathbf{C},\mathbf{D},\mathbf{m},\mathbf{n}) = \lambda (\| \mathbf{A} \|^2 + \| \mathbf{B} \|^2 + \| \mathbf{C} \|^2 + \| \mathbf{D} \|^2 + \| \mathbf{m} \|^2 + \| \mathbf{n} \|^2)$

$\mathrm{S}(\mathrm{x})=\begin{cases}
    \frac{1}{1 + \exp(-\mathrm{x})}, & \text{if x is in a binary column}.\\
    \mathrm{x}, & \text{otherwise}.
  \end{cases}$

$ \mathbf{A}_x = \left(\begin{array}{@{}cc@{}} \mathbf{A}_{\text{shared}} & \mathbf{A}_{\text{main}} \end{array}  \right) , \: \: \mathbf{A}_u = \left(\begin{array}{@{}cc@{}} \mathbf{A}_{\text{attr}} & \mathbf{A}_{\text{shared}} \end{array}  \right)
$

$ \mathbf{B}_x = \left(\begin{array}{@{}cc@{}} \mathbf{B}_{\text{shared}} & \mathbf{B}_{\text{main}} \end{array}  \right) , \: \: \mathbf{B}_i = \left(\begin{array}{@{}cc@{}} \mathbf{B}_{\text{attr}} & \mathbf{B}_{\text{shared}} \end{array}  \right)
$\\

When having non-shared components in the factorizations, the closed-form minimizer for the $\mathbf{A}$ or $\mathbf{B}$ matrices becomes the solution of a linear system with block matrices - for the $\mathbf{A}$ matrix, the solution would be:
$$
\begin{bmatrix} \mathbf{A}_{a} \\ \mathbf{A}_{s} \\ \mathbf{A}_{m} \end{bmatrix}
	=
\begin{pmatrix}
\begin{bmatrix}
\mathbf{0} & \mathbf{C}_{a}^T \\
\mathbf{B}_{s}^T & \mathbf{C}_{s}^T \\
\mathbf{B}_{m} & \mathbf{0}
\end{bmatrix}
\begin{bmatrix}
\mathbf{0} & \mathbf{B}_{s} & \mathbf{B}_{m} \\
\mathbf{C}_{a} & \mathbf{C}_{s} & \mathbf{0}
\end{bmatrix}
+ \text{diag}(\lambda)
\end{pmatrix}^{-1}
\begin{bmatrix}
\mathbf{0} & \mathbf{C}_{a}^T \\
\mathbf{B}_{s}^T & \mathbf{C}_{s}^T \\
\mathbf{B}_{m} & \mathbf{0}
\end{bmatrix}
\begin{bmatrix} \mathbf{X} \\ \mathbf{U} \end{bmatrix}
$$

With the fixed matrices for a given row of $\mathbf{A}$ consisting only of the rows in each respective matrix which are not missing for that user. As such, the model can be fit through the ALS strategy by updating the matrices in sequence, one row at a time, just like in the non-collective case, and in principle this strategy can also be used for fitting variations of this model such as the implicit-feedback variation described in \cite{implicit}.

When introducing sigmoid transformations however, the problem is no longer solvable through a closed-form, but can still be solved using gradient-based methods. The matrices here were optimized using an L-BFGS optimizer (a limited-memory quasi-Newton method \cite{lbfgs}). The implementation of the methods proposed here was made open-source and freely available\footnote{\url{https://github.com/david-cortes/cmfrec}}.

In many implementations of low-rank matrix factorization, the regularization parameters are scaled by the number of ratings by each user and for each movie, but since this model was adding the side information matrices, this idea was not incorporated in the final objective formula.

It can be seen that this optimization objective will produce values for $\mathbf{a}_u$ and $\mathbf{b}_i$ as long as there is either interactions data or side information about a given user $u$ or item $i$. If not applying sigmoid transformations, it is also possible to obtain values for $\mathbf{a}_u$ and $\mathbf{b}_i$ for new users and items based on side information alone without refitting the model entirely using the same closed-form solution as ALS, if holding everything else constant:

$$ \mathbf{a}_u = (\mathbf{C} \mathbf{C}^T + \operatorname{diag}(\lambda))^{-1} \mathbf{C} \mathbf{u}_u $$

and similarly for items:
$$ \mathbf{b}_i = (\mathbf{D} \mathbf{D}^T + \operatorname{diag}(\lambda))^{-1} \mathbf{D} \mathbf{i}_i $$

If using sigmoid or other transformations, such values might still be obtained from solving smaller optimization problems through gradient-based methods. Calculating parameters for new users/items this way, while faster than refitting the entire model from scratch, is still a rather slow process and not fast enough to be used in live systems.

Other approaches similar in spirit have also been proposed, e.g. \cite{zhao}, but they are aimed at warm-start recommendations only.

\section{A different formulation}

Collective matrix factorization as presented here is not the only cold-start-capable model that has been proposed for integrating side information into low-rank matrix factorization models. For example, \cite{offsets} proposed a Bayesian formulation which assumes a decomposable generative model $ \mathbf{X} \approx \mathbf{A}(\mathbf{B} + \mathbf{D}_1), \: \mathbf{I} \approx \mathbf{D}_1 \mathbf{D}_2$ which, informally, can be thought of as calculating a base score for each factor derived from the item attributes, and an offset based on the observed behavior.

While  \cite{offsets} assumed counts data for the attributes and proposed a Bayesian approach to this problem, the idea of decomposing the low-rank matrices into additive components can be used in other settings by following a different optimization route.

As an alternative to the model from the previous section, this work also evaluated a different formulation based on minimizing a loss function as follows:
$$
\min_{\mathbf{A},\mathbf{B},\mathbf{C},\mathbf{D},\mathbf{m},\mathbf{n}}
\| I_x ( \mathbf{X} - \mu - \mathbf{m} - \mathbf{n} - 
(\mathbf{A} + \mathbf{U} \mathbf{C})(\mathbf{B} + \mathbf{I} \mathbf{D})^T  )\|^2
 + \mathrm{R}$$
 
Just like before, optimization was done through L-BFGS. One could also think of solving this problem by first obtaining a solution for $\mathbf{A}$ and $\mathbf{B}$ without the side information, then obtaining $\mathbf{C}$ and $\mathbf{D}$ as the least-squares minimizers for them given by the product of the $\mathbf{U}$ and $\mathbf{I}$ matrices with $\mathbf{C}$ and $\mathbf{D}$ (i.e. solve $\text{min}_{\mathbf{C}} || \mathbf{A}^{*} - \mathbf{U} \mathbf{C} ||^2$, $\mathbf{A} = \mathbf{A}^{*} - \mathbf{U} \mathbf{C}$), but this approach doesn't tend to reach the same final solutions. As well, a gradient-based approach allows other potential enhancements such as setting higher regularization for the free offset.

Informally, this model tries to calculate a base matrix of latent factors by a linear combination of the user/item attributes, to which a free offset is added based on the observed interactions data in order to obtain the final latent factors. It will be referred hereafter as the "offsets" model.

This alternative formulation presents a computational advantage for cold-start recommendations compared to the previous formulation, as now the latent factors based on attributes can be calculated by a simple vector-matrix product $ \mathbf{u}_u \mathbf{C} $ instead of solving a larger linear system, while the offsets are zero in the absence of any interaction data, which makes it suitable for producing cold-start recommendations in real-time. Contrary to the models presented so far, here the low-rank matrices related to the attributes are also used in the prediction formula - predictions for a new user of known items would be given by:
$$ \hat{\mathbf{x}_u} = \mathbf{u}_u \mathbf{C} (\mathbf{B} + \mathbf{I} \mathbf{D})^T + \mathbf{n} $$
(For new items, $\mathbf{b}_i$ and $n_i$ would be zero, while $\mathbf{i}_i \mathbf{D}$ would be calculated in the same way).

It also has the advantages of not requiring any special transformation for variables that are limited in range (e.g. binary or non-negative), and of having fewer hyperparameters to tune.

Compared to other approaches for cold-start recommendations such as \cite{reg} or approches based on user-wise regressions, this model can work in the absence of side information for either users or items, thus being usable in all the different cold-start scenarios, and its parameters are optimized to recommend items based on both attributes and observed interactions.

A further decomposition in which user latent factors are determined separately for combining them with those derived from the interactions data for items and those derived from the item attributes was also explored in \cite{ctpf} ("decoupled" model, $\mathbf{X} \approx \mathbf{A}_1 \mathbf{B} + \mathbf{A}_2 \mathbf{D}_1$) and was briefly attempted here, but the results, in line with \cite{ctpf}, were far below every other model and were left out of the analysis.

\section{Empirical evaluation}

Both of these models were evaluated using the MovieLens 1M dataset \cite{movielens}, complemented with the movie tag genome data \cite{taggenome} taken from the MovieLens latest dataset (last updated 08/2017 at the time these experiments were run), and with demographical and geographical information about users, the later linked to them through their zip code. Unfortunately, later (and larger) releases of the MovieLens dataset no longer include user attributes, nor was the author aware of any larger, public dataset with side information about both users and items, thus it was not possible to evaluate the models on bigger datasets.

The MovieLens 1M dataset contains 1,000,209 ratings on 3,952 movies by 6,040 users, in a timespan from 2000 to 2003. Information from the tag genome dataset, consisting of 1,128 attributes for which movies are assigned a continuous value (which can also be negative) under each of them, was available for 3,028 of these movies only, but demographical information was available for all users, including their age group (7 buckets), occupation (21 categories), gender, and zip code (not used directly), which were taken as binary variables. Additionally, information about the US region of the user (or whether they were not from the US) was added to them as binary variables by linking them through the zip code, using free zip code databases to determine the region\footnote{\url{http://federalgovernmentzipcodes.us/}}\footnote{\url{http://www.fonz.net/blog/archives/2008/04/06/csv-of-states-and-state-abbreviations/}}\footnote{\url{https://www.infoplease.com/us/states/sizing-states}}.

Recommendations were evaluated by randomly splitting the ratings data into a training set and four test sets in order to evaluate the different possible cold-start scenarios and compare them to warm-start recommendations, containing \begin {enumerate*} [label=\itshape\alph*\upshape)] \item only users and items that were in the training data\label{a}, \item users that were not in the training set but items that were\label{b}, \item users that were in the training set but items that were not and had tags available\label{c}, \item users and items that were not in the training set (exact same users as in \ref{b}, and only items that were also in \ref{c})\label{d} \end {enumerate*}, with each test set containing at least 5 ratings from each user included in that set, having sizes as follows:\\

{\centering
\begin{tabular}{|l|c|c|c|}
 \hline
  & \textbf{Ratings} & \textbf{Users} & \textbf{Items} \\
 \hline
\textbf{Train set} & 478,105 & 4,530 & 2,753 \\ \hline
\textbf{Test set 1: users $\in$ train, items $\in$ train} & 81,765 & 3,527 & 2,518 \\ \hline
\textbf{Test set 2: users $\notin$ train, items $\in$ train} & 187,456 & 1,510 & 2,576 \\ \hline
\textbf{Test set 3: users $\in$ train, items $\notin$ train} & 170,579 & 4,278 & 759 \\ \hline
\textbf{Test set 4: users $\notin$ train, items $\notin$ train} & 57,021 & 1,426 & 736 \\
 \hline
\end{tabular}}\\

In the case of the first model, which allows for setting weights for each factorization, different weights were tried to examine their impact on the quality of cold-start and warm-start recommendations. It was also experimented trying to fit the models to side information about users only, items only, or both, and compared to the same model without side informaton. Except when stated otherwise, the models were trained using side information about users and items that had no ratings data - there were tags available for 10,993 movies, many of which were neither in the training nor the test sets.

All the models were fit with the number of latent factors $k = 40$ and regularization $\lambda = 10^{-4}$. The dimensionality of the tags data was reduced by taking only their first 50 principal components, as the number of columns is too large for the second model, and the first model seems to also benefit from reducing dimensionality. Intuitively, however, the first model should not need this type of dimensionality reduction, as it performs it implicitly, but taking advantage of this rich side information would require setting the regularization parameters differently for each matrix, which was not experimented with in here. Some trial and error (not recorded here) suggests that also adding non-shared latent factors brings a slight improvement, particularly when using only item attributes.

Models were evaluated in terms of their RMSE (root mean squared error) and NDCG@5 (net discounted cummulative gain at 5), the later calculated on a per-user basis and averaged across all users. The definition of DCG was taken as follows:
$$ DCG@5 = \sum_{i=1}^{5} \frac{2^{x_i} - 1}{\log_2 (i + 1)} $$
Where $i$ is the item with the $i_{th}$ highest predicted score for a user, and $x$ is the actual rating that the user gave to that item. NDCG is calculated as DCG divided by the maximum possible DCG for that data, resulting in a value that is upper bounded at 1. As this is explicit feedback data, all the evaluation was done on the subset of ratings that were in one of the test sets, rather than comparing predictions for movies that the users rated vs. movies they didn’t rate.

As a baseline, a non-personalized "most-popular" recommendation formula was also evaluated, ranking the items according to their average rating  - it shall be noted that all the items here had a reasonable minimum number of ratings - computed from the ratings in the training set only. Since this simple formula cannot recommend items that were not in the training set, a weaker baseline consisting of random predictions was also evaluated.

{\centering
\begin{table}
\caption {Results on warm-start test set, MovieLens 1M}
\begin{tabular}{|l|c|c|c|c|c|}
 \hline
\textbf{Model} & $\mathbf{w_x}$ & $\mathbf{w_u}$ & $\mathbf{w_i}$ & \textbf{RMSE} & \textbf{NDCG@5} \\ \hline
\textbf{Random recommendations} & - & - & - & - & 0.606 \\ \hline
\textbf{Most-popular} & - & - & - & - & 0.7862 \\ \hline
\textbf{MF, no attributes} & - & - & - & 0.865 & 0.8166 \\ \hline \hline
\textbf{CMF, user attributes*} & 1 & 1 & - & \textbf{0.8635} & 0.8168 \\ \hline
\textbf{CMF, user attributes} & 1 & 1 & - & 0.8638 & \textbf{0.818} \\ \hline
\textbf{CMF, user attributes} & 1.5 & 0.5 & - & 0.8689 & 0.8106 \\ \hline
\textbf{CMF, user attributes} & 0.5 & 1.5 & - & 0.9009 & 0.7977 \\ \hline
\textbf{Offsets, user attributes} & - & - & - & 0.8972 & 0.7968 \\ \hline \hline
\textbf{CMF, item attributes*} & 1 & - & 1 & \textbf{0.8571} & \textbf{0.8207} \\ \hline
\textbf{CMF, item attributes} & 1 & - & 1 & 0.8583 & 0.8201 \\ \hline
\textbf{CMF, item attributes} & 1.5 & - & 0.5 & 0.8616 & 0.8153 \\ \hline
\textbf{CMF, item attributes} & 0.5 & - & 1.5 & 0.8966 & 0.7995 \\ \hline
\textbf{Offsets, item attributes} & - & - & - & 0.9225 & 0.7848 \\ \hline \hline
\textbf{CMF, user and item attributes} & 1 & 1 & 1 & \textbf{0.8575} & \textbf{0.8214} \\ \hline
\textbf{CMF, user and item attributes} & 2.14 & 0.43 & 0.43 & 0.8846 & 0.8029 \\ \hline
\textbf{CMF, user and item attributes} & 0.43 & 1.29 & 1.29 & 0.9030 & 0.7977 \\ \hline
\textbf{Offsets, user and item attributes} & - & - & - & 0.8728 & 0.8072 \\ \hline
\end{tabular}
\begin{tablenotes}
      \small
      \item  * These models were fit to only the attributes data of the users/items with ratings in the same training data
    \end{tablenotes}
\end{table}}

{\centering
\begin{table}
\caption {Results on test set with new users only, MovieLens 1M}
\begin{tabular}{|l|c|c|c|c|c|}
 \hline
\textbf{Model} & $\mathbf{w_x}$ & $\mathbf{w_u}$ & $\mathbf{w_i}$ & \textbf{RMSE} & \textbf{NDCG@5} \\ \hline
\textbf{Random recommendations} & - & - & - & - & 0.5185 \\ \hline
\textbf{Most-popular} & - & - & - & - & 0.7584 \\ \hline \hline
\textbf{CMF, user attributes} & 1 & 1 & - & \textbf{0.9671} & 0.7644 \\ \hline
\textbf{CMF, user attributes} & 1.5 & 0.5 & - & 0.9674 & 0.761 \\ \hline
\textbf{CMF, user attributes} & 0.5 & 1.5 & - & 0.9696 & \textbf{0.7658} \\ \hline
\textbf{Offsets, user attributes} & - & - & - & 0.9804 & 0.7414 \\ \hline \hline
\textbf{CMF, user and item attributes} & 1 & 1 & 1 & 0.9671 & \textbf{0.7638} \\ \hline
\textbf{CMF, user and item attributes} & 2.14 & 0.43 & 0.43 & 0.9670 & 0.7605 \\ \hline
\textbf{CMF, user and item attributes} & 0.43 & 1.29 & 1.29 & 0.9712 & 0.7633 \\ \hline
\textbf{Offsets, user and item attributes} & - & - & - & \textbf{0.9608} & 0.7633 \\ \hline
\end{tabular}\end{table}}

{\centering
\begin{table}
\caption {Results on test set with new items only, MovieLens 1M}
\begin{tabular}{|l|c|c|c|c|c|}
 \hline
\textbf{Model} & $\mathbf{w_x}$ & $\mathbf{w_u}$ & $\mathbf{w_i}$ & \textbf{RMSE} & \textbf{NDCG@5} \\ \hline
\textbf{Random recommendations} & - & - & - & - & 0.5692 \\ \hline \hline
\textbf{CMF, item attributes} & 1 & - & 1 & 1.0128 & 0.7304 \\ \hline
\textbf{CMF, item attributes} & 1.5 & - & 0.5 & 1.0136 & 0.7212 \\ \hline
\textbf{CMF, item attributes} & 0.5 & - & 1.5 & 1.0194 & 0.7302 \\ \hline
\textbf{Offsets, item attributes} & - & - & - & \textbf{0.9558} & \textbf{0.7387} \\ \hline \hline
\textbf{CMF, user and item attributes} & 1 & 1 & 1 & 0.9955 & \textbf{0.7717} \\ \hline
\textbf{CMF, user and item attributes} & 2.14 & 0.43 & 0.43 & 0.9986 & 0.7579 \\ \hline
\textbf{CMF, user and item attributes} & 0.43 & 1.29 & 1.29 & 1.0024 & 0.7657 \\ \hline
\textbf{Offsets, user and item attributes} & - & - & - & \textbf{0.9478} & 0.7642 \\ \hline
\end{tabular}\end{table}}

{\centering
\begin{table}
\caption {Results on test set with new users and new items only, MovieLens 1M}
\begin{tabular}{|l|c|c|c|c|c|}
 \hline
\textbf{Model} & $\mathbf{w_x}$ & $\mathbf{w_u}$ & $\mathbf{w_i}$ & \textbf{RMSE} & \textbf{NDCG@5} \\ \hline
\textbf{Random recommendations} & - & - & - & - & 0.5702 \\ \hline \hline
\textbf{CMF, user and item attributes} & 1 & 1 & 1 & 1.063 & 0.7226 \\ \hline
\textbf{CMF, user and item attributes} & 2.14 & 0.43 & 0.43 & 1.0684 & 0.7239 \\ \hline
\textbf{CMF, user and item attributes} & 0.43 & 1.29 & 1.29 & 1.0598 & \textbf{0.7296} \\ \hline
\textbf{Offsets, user and item attributes} & - & - & - & \textbf{1.0086} & 0.7247 \\ \hline
\end{tabular}\end{table}}

\section{Conclusions and discussion}

This work proposed an enhancement to the collective matrix factorization model in order to deal with binary data, and proposed an alternative formulation - the "offsets" model - that is able to make fast recommendations for new users and items and which does not require any transformation for attributes data that is limited in range.

Cold-start recommendations are understandably not as good as warm-start ones, and the offsets model didn't manage to beat non-personalized recommendations for new users when fit to user side information only, although it is significantly better than random recommendations and it did beat non-personalized recommendations when adding item attributes too. The coverage of these recommendations however is wider than "most-popular" lists, as they can recommend new items when there is information available about them, which in practice can be more valuable than an improvement in offline metrics.

Predictions about new users turned out to be of better quality than predictions about new items, despite side information available about items in this experiment being far more detailed, albeit the difference in the evaluated metrics is not so large. Surprisingly, adding side information about users also resulted in a significant increase in the quality of predictions for new items, moreso than the increase one would expect based on the improvement seen in the warm-start scenario.

In the original formulation, adding side information about users and items that are not in the interactions (ratings) training data seems to worsen warm-start recommendations overall by a very small margin, but a better hyperparameter tuning might be able to make it benefit from this extra information.

Contrary to what one would expect, giving more weight to the factorization of side information in the original model did not result in better cold-start recommendations in most scenarios, nor did giving more weight to the factorization of the main matrix lead to better warm-start recommendations, but rather, the same weights that led to better warm-start recommendations generally also led to better cold-start ones.

Compared to the original CMF, the offsets model resulted in improved cold-start recommendations for new items being recommended to old users when there is no item side information, and depending on the metric being evaluated, also for new items recommended to new users (different runs of all models resulted in slight variations, with the model having highest NDCG@5 not being consistently the same one across runs in the last scenario), but the original CMF model performed slighlty better in recommending old items to new users. Warm-start recommendations from the offsets model however, while generally producing better metrics than non-personalized recommendations, lag behind those from a model without any side information at all.

It should be noted too that the offsets model required 3-4 times more L-BFGS iterations to reach convergence compared to the original model, and in the case of side information for both users and items, was stopped before convergence at 800 iterations. Letting it run for more than 2,000 iterations resulted in significantly degraded results everywhere (not reported here).

Collective matrix factorization models are harder to tune correctly in terms of their hyperparameters when compared with typical matrix factorization models based on interactions alone, and adding more side information with a bad choice of hyperparameters results in worse performance metrics compared to discarding it.

As a remark, this work only evaluated recommendations on one dataset (MovieLens 1M), and it remains to be seen if results will be consistent across different datasets. It also remains to be seen whether the type of attribute data has some influence in the results from each model - the user data here consisted in all-binary columns, while the items data consisted in all-continuous and normally-distributed columns, which might have had an impact in the difference seen between models that take user or item side information.

One serious limitation for both of these models as implemented here is their impracticality in the scenario of implicit data, as running the L-BFGS procedure would require allocating the full user-item interaction matrix, but perhaps the main idea behind the offsets models could also be used with a different optimization procedure, such as \cite{ctpf} or \cite{bpr}.

An interesting possibility that was not explored was to use Bayesian approaches with non-independent latent factors or with some hierarchical structure such as in \cite{gp}, which tend to outperform simpler models based on minimization of squared loss as the ones experimented with in this work \cite{bayes1}\cite{bayes2}.

Another interesting comparison that remains to be seen is against deep-learning-based models such as \cite{nn}, some of which might also result in fast cold-start prediction formulas.

\bibliographystyle{plain}
\bibliography{cmf}

\end{document}